\documentclass[12pt]{article}
\usepackage{amsmath}

\usepackage[super,compress]{cite}
\usepackage{graphicx}
\usepackage[polutonikogreek,english]{babel}
\usepackage{comment}
\renewcommand{\d}{\mathrm{d}}
\newcommand{\bpart}{{\mbox{\boldmath$\partial$}}}
\newcommand{\ralpha}{{\rm \textgreek{a}}}
\newcommand{\rlambda}{{\rm \textgreek{l}}}
\newcommand{\rnu}{{\rm \textgreek{n}}}
\newcommand{\rkappa}{{\rm \textgreek{k}}}
\newcommand{\rmu}{{\rm \textgreek{m}}}
\newcommand{\trmu}{{\widetilde{\rmu}}}
\newcommand{\on}{{\overline{n}}}
\newcommand{\otheta}{\overline{\vartheta}}
\newcommand{\Tr}{\mathop{\mathrm{Tr}}}
\newcommand{\Det}{\mathop{\mathrm{Det}}}
\newcommand{\bp}{{\mbox{\boldmath$p$}}}

\renewcommand{\Re}{\mathop{\rm Re}\nolimits}
\renewcommand{\Im}{\mathop{\rm Im}\nolimits}

\allowdisplaybreaks[4]

\begin{document}

\title{Soft synchronous gauge: principal value prescription}

\author{V.M. Khatsymovsky \\
 {\em Budker Institute of Nuclear Physics} \\ {\em of Siberian Branch Russian Academy of Sciences} \\ {\em
 Novosibirsk,
 630090,
 Russia}
\\ {\em E-mail address: khatsym@gmail.com}}
\date{}
\maketitle

\begin{abstract}

The synchronous gauge in gravity ($g_{0 \lambda} = - \delta_{0 \lambda}$) is ill-defined due to the singularity at $p_0 = 0$ in the graviton propagator. Previously we studied "softening" this gauge by considering instead the gauge $n^\lambda g_{\lambda \mu} = 0$, $n^\lambda = (1, - \varepsilon (\partial^j \partial_j )^{- 1} \partial^k ) $ in the limit $\varepsilon \to 0$. We now explore the possibility of using a principal value prescription (not in the standard Cauchy sense), which amounts, roughly speaking, to replacing singularities $p_0^{-j} \Rightarrow [ (p_0 + i \varepsilon )^{-j} + (p_0 - i \varepsilon )^{-j} ] / 2$, which then behave like distributions. We show that such a propagator follows upon adding to the action a gauge-violating term of a general form, which reduces to $ \sim \int f_\lambda \Lambda^{\lambda \mu} f_\mu \d^4 x $ with a constant operator $\Lambda^{\lambda \mu}$ depending on $\partial$ and a metric functional $f_\lambda$. The contribution of the ghost fields to the effective action is analysed. For the required intermediate regularization, the discrete structure of the theory at small distances is implied. It is shown that the ghost contribution can be disregarded in the limit $ \varepsilon \to 0$.

\end{abstract}

PACS Nos.: 04.60.-m

MSC classes: 83C45; 83C47

keywords: general relativity; Feynman diagrams; synchronous gauge; functional integral; temporal gauge

\section{Introduction}

General relativity (GR) at the perturbative level is a non-renormalizable theory \cite{hooft}. However, we can regard it as an effective low-energy theory \cite{don,don1,don2,don3}, in which the properties of the underlying theory have an imprint in the form a number of phenomenological constants. For example, in a number of works \cite{don,don1,don2,don3,muz,akh,hamliu1,kk,kk1} long-range quantum corrections to the Newtonian potential caused by the exchange of gravitons were studied.

We are interested in the synchronous gauge $g_{0 \lambda} = \eta_{0 \lambda}$, here in the aspect of Feynman diagrams, where $\eta_{\lambda \mu} = {\rm diag} (-1, 1, 1, 1)$. An advantage of such a gauge is that we are dealing only with physically meaningful metric variables, which here are the six spatial metric components. Using synchronous coordinates can often be useful, including at the nonperturbative level. For example, in \cite{Frob} it is proposed to define observables in the dynamical synchronous coordinates and it is shown that these observables are gauge-independent. Along this way, it may be of interest to express objects such as the graviton propagator in synchronous coordinates. To analyze the possibility of implementing the synchronous gauge condition in the asymptotic safety program for gravity may also be of interest; there the harmonic coordinate condition is mostly employed \cite{Reut}. This approach provides a nonperturbative notion of renormalization.

The disadvantage of the synchronous gauge is the appearance of poles at $p_0 = 0$ (up to $p_0^{- 4}$) in the graviton propagator. This is a consequence of the fact that the condition $g_{0 \lambda} = \eta_{0 \lambda}$ fixes the gauge only up to arbitrary time-independent gauge transformations. The synchronous gauge is analogous to the $nA^a = 0$ gauge in Yang-Mills theories, where typically the constant 4-vector $n$ is equal to $(1,0,0,0)$ (time gauge) or $(0,0,0,1)$ (axial gauge)), which has been used for quite a long time \cite{Schw}. An advantage of this gauge is the absence of the ghost field contribution, but the disadvantage is again the appearance of a (double) pole at $np = 0$ in the gauge field propagator. To correct this disadvantage, Landshoff has proposed a prescription \cite{Land} that involves the replacement $1/p_0^2 \Rightarrow 1/(p_0^2 + \varepsilon^2), \varepsilon \to 0$ in the propagator (if $n = (1,0,0,0)$). Steiner has justified this prescription by"softening" the gauge condition $A^a_0 = 0$ as follows \cite{Ste}:
\begin{equation}\label{nA=0}                                                %1
n^\lambda A^a_\lambda = 0, ~ n^\lambda = \rnu^\lambda - \varepsilon \frac{ \partial_{\! \! \perp}^\lambda}{ \partial_{\! \! \perp}^2 }, ~ \partial_{\! \! \perp \lambda} = \partial_\lambda - \rnu_\lambda \frac{\rnu \partial}{ \rnu^2 } , ~ \rnu^\lambda = (1, 0, 0, 0) , ~ \partial_{\! \! \perp \lambda} = (0, \bpart) .
\end{equation}

\noindent The 4-vector $\rnu^\lambda$ indicates the temporal direction, $\partial_{\! \! \perp \lambda}$ is the transverse (spatial) derivative. That is, $n^\lambda$ becomes a differential nonlocal operator, and nonphysical singularities are shifted into the plain of complex momentum, $( n p )^{- 1} = (p_0 + i \varepsilon )^{- 1}$, $( \on p )^{- 1} = (p_0 - i \varepsilon )^{- 1}$.

In the soft synchronous gauge in gravity, $n^\lambda g_{\lambda \mu} = 0$, we obtain an analogue of the Landshoff prescription and find \cite{khat1} that the contribution of the ghost field to the effective action (in fact, the contribution to the functional integral measure, since the contribution to the effective action turns out to be imaginary) vanishes in the limit $\varepsilon \to 0$. It turns out to be important that the effective ghost action is of order $O( \varepsilon^2 )$, since then we can correctly organize the transition to the limit $\varepsilon \to 0$ and disregard this action.

Now we would like to approach this gauge so as to have for the singularities in the propagator their replacement of the type $p_0^{-j} \Rightarrow [ (p_0 + i \varepsilon )^{-j} + (p_0 - i \varepsilon )^{-j} ] / 2$, which gives $p_0^{-j}$ in the limit $\varepsilon \to 0$ the properties of a generalized function with the usual differential relations like $(\partial / \partial p_0 ) p_0^{-j} = - j p_0^{-j-1}$. Also in a Feynman diagram this leads to integrals of the type
\begin{equation}\label{int-(p0+ie)^(-j)...+(+to-)}                         %2
\int \frac{ \d p_0 }{ 2 \pi } \frac{1}{(p_0 + i \varepsilon )^j} \cdot \dots \mbox{ and } \int \frac{ \d p_0 }{ 2 \pi } \frac{1}{(p_0 - i \varepsilon )^j} \cdot \dots .
\end{equation}

\noindent Here the dots mean the rest of the diagram, the poles of which in $p_0$ for $\varepsilon \to 0$ lie at $p_0 = O(1)$ (except for a set of zero measure in the space of other loop and external momenta of the diagram), since generally the momenta of the internal lines are different. Therefore, we can deform the integration contour in the plane of complex $p_0$ so that it would pass at distances $O(1)$ from all singularities, and the result of integration over $\d p_0$ will be $O(1)$. This is a pleasant feature compared to the above prescription $1/p_0^2 \Rightarrow 1/(p_0^2 + \varepsilon^2) = (p_0 + i \varepsilon )^{-1} (p_0 - i \varepsilon )^{-1}$, where the integration contour is clamped between $p_0 = i \varepsilon$ and $p_0 = - i \varepsilon$. Of course, when summing different diagrammatic contributions and/or omitting non-pole terms that do not contribute to the absorptive part of the S-matrix, a well-defined contribution follows, but it appears to be an advantage of the considered prescription that the various diagrammatic contributions are initially finite individually.

A procedure of this type, called principal value prescription (but not in the Cauchy sense), can be implemented in the Yang-Mills case by preserving the gauge-breaking term $ \sim \int ( n^\lambda A^a_\lambda ) \rlambda ( n^\mu A^a_\mu ) \d^4 x $ with $\rlambda^{- 1} \neq 0$, choosing $\rlambda$ and $\varepsilon$ as functions of $\partial$ themselves (in particular, $\varepsilon$ is substituted by $\varepsilon^2 / ( \rnu \partial )$), and taking the limit $\varepsilon \to 0$, $\rlambda \to \infty$ \cite{Vel}. The choice of two functions of $\partial$ ($\rlambda$ and $\varepsilon$) turns out to be sufficient to soften in the desired way the two functions $p_0^{- 1}$ and $p_0^{- 2}$ appearing in the propagator. In gravity, the adoption of a gauge-violating term in the form $ \sim \int (n^\mu w_{\mu \lambda}) \rlambda^{\lambda \sigma } (n^\tau w_{\tau \sigma}) \d^4 x $, $w_{\lambda \mu} = g_{\lambda \mu} - \eta_{\lambda \mu}$, and the accepted freedom in choosing $\rlambda^{\lambda \sigma }$ and $\varepsilon$ as functions of $\partial$ does not allow all four functions $p_0^{-j}$, $j = 1 - 4$, to be softened, since the same $p_0^{-j}$, multiplying different tensorial structures in the propagator, is softened (by selecting the dependence of $ \rlambda^{\lambda \mu} $ and $ \varepsilon $ on $\partial$) differently, and for some structures is not softened at all. Instead, we can consider a half-sum of non-Hermitian but mutually Hermitian adjoint operators ("artificial" propagators), each of which is the propagator expression, in one of which both $n, \on$ are replaced by $n$, in the other by $\on$. (This can be thought of as a matrix analogue of the principal value.) We can then find the corresponding gauge-violating term. A priori, this is a general bilinear form on $w_{\lambda \mu}$, which over symmetric pairs of vector indices is a general $10 \times 10$ matrix. But the calculation gives for this term an expression of the form $ \sim \int ( {\rm O }^{\lambda \mu}_\rho w_{\lambda \mu} ) \Lambda^{\rho \kappa} ( {\rm O }^{\sigma \tau}_\kappa w_{\sigma \tau} ) \d^4 x $, where the operators ${\rm O }^{\lambda \mu}_\rho$, $\Lambda^{\rho \kappa}$ are functions of $\partial$. Then we can analyze the contribution of the ghost fields to the effective action (the result of functional integration over ghosts).

It turns out that the effective ghost action is $O( \varepsilon^2 )$. Therefore, as said above, we can correctly organize the transition to the limit $\varepsilon \to 0$ and disregard this action. Or, in other words (since the effective ghost action is imaginary), the functional integral measure is not modified due to the ghost field in this limit.

Soft synchronous gauge and the ghost effect on the functional measure may be of interest in connection with the above-mentioned treatment of GR as an effective low-energy theory and the adoption of the underlying theory as discrete. In the Feynman diagrammatic context, elements of the appropriate diagram technique should follow from studying the further specified underlying theory. The latter may be simplicial, based on the Regge calculus \cite{Regge}. It is a closed theory that can be quantized and make predictions for physical effects/constants \cite{RocWilPL,RocWil,HamWil1}. The usual continuum symmetries must be restored by summing/averaging over all possible simplicial structures in the functional integral. We find \cite{our2} that the Regge action can be reduced in the leading order in metric variations to a finite-difference form of the continuum Hilbert-Einstein action. Functional integration over a discrete connection variable in the connection representation of the Regge action yields some functional measure that singles out certain spacetime scales of the metric/length variables proportional to the Planck scale (the appearance of these scales is a non-perturbative effect). The usual requirement of an extremum of the action together with the requirement of the maximum of the measure (which gives typical elementary length scales) gives the initial point of the perturbative expansion. It is in the vicinity of this point that we formulate the perturbative expansion of interest to us (UV finite) \cite{our1,khat}. This expansion consists of discrete analogues of all standard continuum diagrams and diagrams with some new vertices. We can exactly integrate functionally over a discrete connection and obtain a functional measure on the particular set of configurations with an arbitrarily small elementary length scale in the temporal direction. We show \cite{khat} that for the perturbative expansion at such a point (configuration) to be free from powers of a large parameter, length variations in the temporal direction or variations in the discrete ADM lapse-shift functions \cite{ADM1} must be forbidden. Thus we come to the synchronous gauge. Previously we considered the graviton propagator in such a gauge \cite{our1} with its inherent shortcomings, and the present “soft” synchronous gauge aims to eliminate them.

In Section \ref{graviton} we find the propagator for the gauge-violating term $ \sim \int (n^\mu w_{\mu \lambda}) \rlambda^{\lambda \sigma }$ $\cdot (n^\tau w_{\tau \sigma}) \d^4 x $ with the general matrix $\| \rlambda^{\lambda \sigma} \|$, write out the corresponding principal value propagator, and find the parity-violating term that gives the principal value propagator. In Section \ref{ghost} we write out the operator ${\cal O}$ whose determinant is the ghost field determinant, analyze the various contributions to the effective ghost action $S_{\rm ghost} = - i \ln \Det {\cal O}$, and find that it is of order $O ( \varepsilon^2 )$. This means, as said above, that the ghost contribution to the functional integral measure can be disregarded in the limit $\varepsilon \to 0$.

\section{Graviton propagator and the gauge-violating term}\label{graviton}

We issue from the graviton propagator of possibly general form, corresponding to fixing $n^\mu w_{\mu \lambda}$; $w_{\lambda \mu} = g_{\lambda \mu} - \eta_{\lambda \mu}$, $n^\mu$ being a 4-vector (maybe even a differential operator, but not a function of coordinates), in order to form from it an object of the principal value type. We proceed from the Einstein-Hilbert action
\begin{eqnarray}                                                            %3
S & = & \frac{1}{8} \int \d^4 x \sqrt{ - g } g_{\lambda \mu , \nu} g_{\rho \sigma , \tau} \left( 2 g^{\lambda \rho} g^{\mu \tau} g^{\nu \sigma} - g^{\lambda \rho} g^{\mu \sigma} g^{\nu \tau} - 2 g^{\lambda \tau} g^{\mu \nu} g^{\rho \sigma} \right. \nonumber \\ & & \left. + g^{\lambda \mu} g^{\rho \sigma} g^{\nu \tau} \right)
\end{eqnarray}

\noindent (in Planck units; or, in ordinary units, this must be multiplied by $(8 \pi G)^{- 1}$, where $G$ is Newton's constant here) and that one with the source term and a general bilinear in $n^\mu w_{\mu \lambda}$ gauge-fixing term,
\begin{eqnarray}\label{S[J]}                                                %4
S^\prime [ J ] & = & S - \int \d^4 x \left[ J^{\lambda \mu} w_{\lambda \mu } + \frac{1}{4} (n^\mu w_{\mu \lambda}) \rlambda^{\lambda \sigma } (n^\tau w_{\tau \sigma}) \right] , \quad g_{\lambda \mu} = \eta_{\lambda \mu} + w_{\lambda \mu} , \nonumber \\ \eta^{\lambda \mu } & = & {\rm diag} (-1, 1, 1, 1) , ~~~ ( \| \rlambda^{\lambda \mu} \|^{- 1} )_{\sigma \tau} \equiv \ralpha_{\sigma \tau} , \mbox{ ~ for $n^\lambda$ see (\ref{nA=0}), }
\end{eqnarray}

Notations: For the matrices $\rlambda$, $\ralpha$, the 4-vector $\rnu$, and below the 4-vector field $\rkappa$ and the coefficients $\rmu$ we use upright letters to distinguish them from the indices $\lambda$, $\alpha$, $\nu$, $\kappa$, $\mu$ which are written in italics.

To determine the graviton propagator, we vary $S^\prime [ J ]$ with respect to $w_{\lambda \mu}$. The derivation generalizes that one for the case $\rlambda^{\lambda \sigma } = \rlambda \eta^{\lambda \sigma }$ from our preceding paper \cite{khat1}. Equating the result of the variation to zero gives
\begin{eqnarray}\label{d2g=J+nf+d2g}                                        %5
\partial^2 w_{\lambda \mu} & = & 4 J_{\lambda \mu} + \on_\mu \rlambda_\lambda{}^\rho w_{\rho \nu} n^\nu + \on_\lambda \rlambda_\mu{}^\rho w_{\rho \nu} n^\nu + \partial_\mu \partial^\nu w_{\lambda \nu} + \partial_\lambda \partial^\nu w_{\mu \nu} \nonumber \\ & & - \eta^{\nu \rho} \partial_\lambda \partial_\mu w_{\nu \rho} - \eta_{\lambda \mu} \partial^\nu \partial^\rho w_{\nu \rho} + \eta_{\lambda \mu} \eta^{\nu \rho} \partial^2 w_{\nu \rho} .
\end{eqnarray}

\noindent We denote $F^\lambda = \rlambda^{\lambda \rho} w_{\rho \nu} n^\nu$, $f_\lambda = n^\mu w_{\lambda \mu} = \ralpha_{\lambda \mu} F^\mu$. Taking the divergence of both sides using the operator $\partial^\mu ( \cdot )$, we get
\begin{equation}                                                            %6
\on_\mu \partial^\mu F_\lambda + \on_\lambda \partial^\mu F_\mu = - 4 \partial^\mu J_{\lambda \mu} .
\end{equation}

\noindent This gives $F_\lambda$,
\begin{equation}                                                            %7
F_\lambda = - 4 (\on \partial )^{- 1} \partial^\mu J_{\lambda \mu} + 2 \on_\lambda (\on \partial )^{- 2} \partial^\mu \partial^\nu J_{\mu \nu} .
\end{equation}

\noindent Taking the trace of (\ref{d2g=J+nf+d2g}) gives $\eta^{\lambda \mu} \partial^2 w_{\lambda \mu} - \partial^\lambda \partial^\mu w_{\lambda \mu}$ in terms of $J_{\lambda \mu}$ and the found $F_\lambda$,
\begin{equation}                                                           %8
\eta^{\lambda \mu} \partial^2 w_{\lambda \mu} - \partial^\lambda \partial^\mu w_{\lambda \mu} = - 2 \eta^{\lambda \mu} J_{\lambda \mu} - \on^\lambda F_\lambda .
\end{equation}

\noindent Multiplying (\ref{d2g=J+nf+d2g}) by $n^\mu$ allows to express $\partial^\mu w_{\lambda \mu} - \eta^{\mu \nu} \partial_\lambda w_{\mu \nu}$ in terms of $J_{\lambda \mu}$ and the found $F_\lambda$, $f_\lambda = \ralpha_{\lambda \mu} F^\mu$ and $\eta^{\lambda \mu} \partial^2 w_{\lambda \mu} - \partial^\lambda \partial^\mu w_{\lambda \mu}$,
\begin{eqnarray}                                                            %9
( n \partial ) ( \partial^\mu w_{\lambda \mu} - \eta^{\mu \nu} \partial_\lambda w_{\mu \nu} ) & = & - 4 n^\mu J_{\lambda \mu} + \partial^2 f_\lambda - \partial_\lambda \partial^\mu f_\mu - ( n \on ) F_\lambda - \on_\lambda n^\mu F_\mu \nonumber \\ & & - n_\lambda ( \eta^{\mu \nu} \partial^2 w_{\mu \nu} - \partial^\mu \partial^\nu w_{\mu \nu} ) .
\end{eqnarray}

\noindent In turn, multiplying the found $\partial^\mu w_{\lambda \mu} - \eta^{\mu \nu} \partial_\lambda w_{\mu \nu}$ by $n^\lambda$ and using the found $f_\lambda$, $F_\lambda$ gives $\eta^{\lambda \mu} w_{\lambda \mu}$. Substituting the found $\eta^{\lambda \mu} w_{\lambda \mu}$ back into $\partial^\mu w_{\lambda \mu} - \eta^{\mu \nu} \partial_\lambda w_{\mu \nu}$, we find $\partial^\mu w_{\lambda \mu}$. Thus, the terms with $w_{\lambda \mu}$ appearing on the RHS of (\ref{d2g=J+nf+d2g}) have been found. Substituting them there, we obtain the propagator, $w_{\lambda \mu} = G_{\lambda \mu \sigma \tau} J^{\sigma \tau}$,
\begin{eqnarray}\label{G}                                                  %10
G_{\lambda \mu \sigma \tau} ( n, \on ) & = & - i \langle w_{\lambda \mu} w_{\sigma \tau} \rangle = \frac{2}{\partial^2 } [ L_{\lambda \sigma} ( n, \on ) L_{\mu \tau} ( n, \on ) + L_{\mu \sigma} ( n, \on ) L_{\lambda \tau} ( n, \on ) \nonumber \\ & - & L_{\lambda \mu} ( n, n ) L_{\sigma \tau} ( \on, \on ) ] - 2 \frac{ ( \ralpha_{\lambda \sigma} \partial_\tau + \ralpha_{\lambda \tau} \partial_\sigma) \partial_\mu + ( \ralpha_{\mu \sigma} \partial_\tau + \ralpha_{\mu \tau} \partial_\sigma) \partial_\lambda }{(n \partial)(\on \partial)} \nonumber \\ & + & 2 \left[ \partial_\lambda \partial_\mu \frac{ n^\nu \ralpha_{\nu \sigma} \partial_\tau + n^\nu \ralpha_{\nu \tau} \partial_\sigma }{(n \partial)^2 (\on \partial)} + \frac{ \ralpha_{\lambda \nu} \on^\nu \partial_\mu + \ralpha_{\mu \nu} \on^\nu \partial_\lambda }{(n \partial) (\on \partial)^2 } \partial_\sigma \partial_\tau \right] \nonumber \\ & - & 2 \frac{ n^\nu \ralpha_{\nu \rho} \on^\rho }{(n \partial)^2 (\on \partial)^2 } \partial_\lambda \partial_\mu \partial_\sigma \partial_\tau .
\end{eqnarray}

\noindent Here
\begin{equation}                                                           %11
L_{\lambda \mu} ( m, n ) \stackrel{\rm def }{=} \eta_{\lambda \mu} - \partial_\lambda \frac{m_\mu }{ m \partial } - \frac{n_\lambda }{ n \partial } \partial_\mu + \frac{(m n) \partial_\lambda \partial_\mu}{(m \partial )( n \partial )}
\end{equation}

\noindent for 4-vectors $m$, $n$.

In the momentum representation and for $\rnu^\lambda = (1, 0, 0, 0)$, we have $-in \partial = p_0 + i \varepsilon$, $-i\on \partial = p_0 - i \varepsilon$. The principal value prescription for any (negative) power of $p_0$ is the half-sum of this power of $p_0 + i \varepsilon$ and of $p_0 - i \varepsilon$. Not to be confused with the principal value in Cauchy's sense which means nullifying this function in the interval $(- \varepsilon, \varepsilon)$ in the limit $\varepsilon \to 0$ when integrating it. Thus, for the propagator we would like to have
\begin{equation}\label{vpG=}                                               %12
G_{\lambda \mu \sigma \tau} = \frac{1}{2} G_{\lambda \mu \sigma \tau} ( n, n ) + \frac{1}{2} G_{\lambda \mu \sigma \tau} ( \on, \on ) .
\end{equation}

\noindent This does not exactly correspond to performing exclusively the operation $p_0^{-j} \Rightarrow [ (p_0 + i \varepsilon )^{-j} + (p_0 - i \varepsilon )^{-j} ] / 2$, $j = 1, 2, 3, 4$, on $p_0^{-j}$ included in $G_{\lambda \mu \sigma \tau} ( \rnu, \rnu )$. However, the key property that each term has poles only in the upper or only in the lower half-plane of the complex $p_0$ is still preserved, which is essential for good calculation properties of such a prescription, as mentioned in the paragraph with (\ref{int-(p0+ie)^(-j)...+(+to-)}). This is a half-sum of non-Hermitian operators, which can formally be viewed as inverses of the non-Hermitian matrices
\begin{equation}\label{M-nn,M-n*n*}                                        %13
{\cal M}^{\lambda \mu \sigma \tau} - \frac{1}{2} n^{( \lambda} \rlambda^{\mu )( \sigma} n^{ \tau )}, ~~~ {\cal M}^{\lambda \mu \sigma \tau} - \frac{1}{2} \on^{( \lambda} \rlambda^{\mu )( \sigma} \on^{ \tau )} ,
\end{equation}

\noindent where ${\cal M }^{\lambda \mu \nu \rho}$ is the bilinear form of the original gravity action,
\begin{equation}                                                           %14
S = \int \d^4 x \frac{1}{2} w_{\lambda \mu} {\cal M}^{\lambda \mu \sigma \tau} w_{\sigma \tau} + O ( w^3 ) .
\end{equation}

\noindent We can write these matrices as sums of a Hermitian one ${\cal C}$ (their $\Re$-part in the momentum representation) and anti-Hermitian ones $\pm i {\cal E}$ (their $i \Im$-parts in the momentum representation) for some Hermitian ${\cal C}$ and ${\cal E}$,
\begin{eqnarray}                                                           %15
{\cal C} = \left \| {\cal C}^{(\lambda \mu)(\sigma \tau)} \right \| = \left \| {\cal M}^{\lambda \mu \sigma \tau} - \frac{1}{2} \Re n^{( \lambda} \rlambda^{\mu )( \sigma} n^{ \tau )} \right \| , \nonumber \\ {\cal E} = \left \| {\cal E}^{(\lambda \mu)(\sigma \tau)} \right \| = \left \| - \frac{1}{2} \Im n^{( \lambda} \rlambda^{\mu )( \sigma} n^{ \tau )} \right \|
\end{eqnarray}

Note that a direct variation of the quadratic forms described by the non-Hermitian matrices (\ref{M-nn,M-n*n*}) under the integral sign in action will lead to their Hermitian symmetrization due to integration by parts and to Hermitian propagators, and not to the non-Hermitian propagators $G ( n, n )$, $G ( \on, \on )$ that interest us, but formally we can represent these propagators as inverses of such forms; also these propagators can be viewed as analytic continuations of the propagator $G ( \rnu, \rnu )$ for $\varepsilon = 0$ ("hard") to complex $\rnu = n$ or $\rnu = \on$.

We aim at $\ralpha_{\lambda \mu} \to 0$, $\varepsilon \to 0$, but now we can not set from the very beginning $\ralpha_{\lambda \mu} = 0$, since the result should be an analytical continuation of the gauge $\rnu^\lambda w_{\lambda \mu} = 0$ to complex $\rnu = n$ or $\rnu = \on$. But this gauge is described by the factor $\delta^4 ( \rnu^\lambda w_{\lambda \mu} )$ in the functional integral, and the $\delta$-function of the complex argument is not defined. Thus, the effects of nonzero $\ralpha_{\lambda \mu}$, $\varepsilon$ should be comparable. The magnitude of these effects is manifested in a nonzero value of $\rnu^\lambda w_{\lambda \mu}$ for the metrics contributing to the propagators, $G ( n, n )$ and $G ( \on, \on )$, considered as correlators. Namely, $n^\lambda w_{\lambda \mu}$ or $\on^\lambda w_{\lambda \mu}$ are $O( 1 / \sqrt{ | \rlambda | } ) = O( \sqrt{ | \ralpha | })$ for these metrics, and in general $\rnu^\lambda w_{\lambda \mu} = O( 1 / \sqrt{ | \rlambda | } ) + O( \varepsilon ) = O( \sqrt{ | \ralpha | }) + O( \varepsilon )$. Here $\ralpha$, $\rlambda$ are typical scales of $\ralpha_{\lambda \mu}$, $\rlambda^{\lambda \mu}$. Thus, $\sqrt{ | \ralpha | }$ and $\varepsilon$ should be comparable as they approach zero, that is, $\ralpha_{\lambda \mu} = O( \varepsilon^2 )$.

The definition of the principal value propagator (\ref{vpG=}) takes the form
\begin{equation}                                                           %16
G = \frac{1}{2} ( {\cal C } + i {\cal E } )^{-1} + \frac{1}{2} ( {\cal C } - i {\cal E } )^{-1} = ( { \cal C } + { \cal E } { \cal C }^{-1} { \cal E } )^{-1} .
\end{equation}

\noindent This is a matrix analog of the principal value
\begin{equation}                                                           %17
\frac{1}{2} \left ( \frac{1}{c + i \varepsilon } + \frac{1}{c - i \varepsilon } \right ) = \frac{ c }{ c^2 + \varepsilon^2 } = \frac{1 }{ c + \varepsilon^2 / c }
\end{equation}

\noindent of $c^{-1}$ for c-numbers $c$ and $\varepsilon \to 0$.

Thus, the propagator of interest is the inverse of ${ \cal C } + { \cal E } { \cal C }^{-1} { \cal E }$, which is a matrix ${\cal M }^{\lambda \mu \sigma \tau} + \Delta {\cal M }^{\lambda \mu \sigma \tau}$, where the additional term is
\begin{eqnarray}                                                           %18
\Delta {\cal M }^{\lambda \mu \sigma \tau} & = & - \frac{1}{2} \Re \left( n^{( \mu } \rlambda^{\lambda ) ( \sigma } n^{\tau )}\right) + \frac{1}{4} \left[ \Im \left( n^{( \mu } \rlambda^{\lambda ) \zeta } n^{\pi }\right) \right] G^{\rm aux}_{\zeta \pi \chi \psi} \left[ \Im \left( n^{ \chi } \rlambda^{\psi ( \sigma } n^{\tau )}\right) \right], \nonumber \\ G^{\rm aux}_{\zeta \pi \chi \psi} & = & \left[ \left\| {\cal M }^{\nu \rho \kappa \varphi} - \frac{1}{2} \Re \left( n^{( \nu } \rlambda^{\rho ) ( \kappa } n^{\varphi )}\right) \right\|^{-1} \right]_{\zeta \pi \chi \psi}.
\end{eqnarray}

\noindent Here
\begin{equation}\label{Re(n-lambda-n)}                                     %19
\Re \left( n^{ \nu } \rlambda^{\rho \kappa } n^{\varphi }\right) = \rnu^{ \nu } \rlambda^{\rho \kappa } \rnu^{\varphi } + \varepsilon^2 \frac{ \partial_{\! \! \perp}^\nu }{ \partial_{\! \! \perp}^2 } \rlambda^{\rho \kappa } \frac{ \partial_{\! \! \perp}^\varphi }{ \partial_{\! \! \perp}^2 }.
\end{equation}

We denote
\begin{equation}                                                           %20
G^{(0)}_{\zeta \pi \chi \psi} = \left[ \left\| {\cal M }^{\nu \rho \kappa \varphi} - \frac{1}{2} \rnu^{( \nu } \rlambda^{\rho ) ( \kappa } \rnu^{\varphi )} \right\|^{-1} \right]_{\zeta \pi \chi \psi}
\end{equation}

\noindent (the usual "hard" synchronous gauge propagator) and express the auxiliary propagator $G^{\rm aux}_{\zeta \pi \chi \psi}$ in terms of $G^{(0)}_{\zeta \pi \chi \psi}$ and the additional $\varepsilon^2$-term in (\ref{Re(n-lambda-n)}),
\begin{equation}                                                           %21
G^{\rm aux}_{\zeta \pi \chi \psi} = G^{(0)}_{\zeta \pi \chi \psi} + \frac{\varepsilon^2}{2} G^{(0)}_{\zeta \pi \nu \rho } \frac{ \partial_{\! \! \perp}^\nu }{ \partial_{\! \! \perp}^2 } \left[ \left\| \ralpha_{\lambda \tau} - \frac{\varepsilon^2}{2} \frac{ \partial_{\! \! \perp}^\mu }{ \partial_{\! \! \perp}^2 } G^{(0)}_{\lambda \mu \sigma \tau} \frac{ \partial_{\! \! \perp}^\sigma }{ \partial_{\! \! \perp}^2 } \right\|^{ - 1 } \right]^{ \rho \kappa } \frac{ \partial_{\! \! \perp}^\varphi }{ \partial_{\! \! \perp}^2 } G^{(0)}_{\kappa \varphi \chi \psi}.
\end{equation}

Also denote
\begin{equation}\label{a-dGd=M}                                            %22
\frac{(\rnu \partial )^2}{\varepsilon^2 } \ralpha_{\lambda \tau} - \frac{(\rnu \partial )^2 }{2} \frac{ \partial_{\! \! \perp}^\mu }{ \partial_{\! \! \perp}^2 } G^{(0)}_{\lambda \mu \sigma \tau} \frac{ \partial_{\! \! \perp}^\sigma }{ \partial_{\! \! \perp}^2 } = M_{\lambda \tau}.
\end{equation}

\noindent (Due to the above accepted relation $\ralpha_{\lambda \mu} = O( \varepsilon^2 )$, the value of $M_{\lambda \tau}$ is of order $O ( 1 )$ when passing to the limit $\varepsilon \to 0$.) Then
\begin{eqnarray}                                                           %23
\Delta {\cal M }^{\lambda \mu \sigma \tau} & = & - \frac{1}{2} \Re \left( n^{( \mu } \rlambda^{\lambda ) ( \sigma } n^{\tau )}\right) - \frac{1}{4} \left[ i \Im \left( n^{( \mu } \rlambda^{\lambda ) \zeta } n^{\pi }\right) \right] \left[ \vphantom{ \frac{ \partial_{\! \! \perp}^\nu }{ \partial_{\! \! \perp}^2 } } G^{(0)}_{\zeta \pi \chi \psi} \right. \nonumber \\ & + & \left. \frac{(\rnu \partial )^2}{2} G^{(0)}_{\zeta \pi \nu \rho } \frac{ \partial_{\! \! \perp}^\nu }{ \partial_{\! \! \perp}^2 } ( M^{- 1} )^{ \rho \kappa } \frac{ \partial_{\! \! \perp}^\varphi }{ \partial_{\! \! \perp}^2 } G^{(0)}_{\kappa \varphi \chi \psi} \right] \left[ i \Im \left( n^{ \chi } \rlambda^{\psi ( \sigma } n^{\tau )}\right) \right].
\end{eqnarray}

Then we further transform this expression. According to
\begin{eqnarray}\label{Im(nn)Im(nn)}                                       %24
& & \left[ i \Im ( n \otimes n ) \right] \otimes \left[ i \Im ( n \otimes n ) \right] = \frac{ \epsilon^2 }{ \left( \partial_{\! \! \perp}^2 \right)^2 } \left( \rnu \otimes \partial_{\! \! \perp} + \partial_{\! \! \perp} \otimes \rnu \right) \otimes \left( \rnu \otimes \partial_{\! \! \perp} + \partial_{\! \! \perp} \otimes \rnu \right) = \frac{ \epsilon^2 }{ \left( \partial_{\! \! \perp}^2 \right)^2 } \nonumber \\ & & \cdot \left\{ \rnu \otimes \partial_{\! \! \perp} \otimes \partial_{\! \! \perp} \otimes \rnu + \left( \partial_{\! \! \perp} \otimes \rnu \otimes \partial_{\! \! \perp} \otimes \rnu + \rnu \otimes \partial_{\! \! \perp} \otimes \rnu \otimes \partial_{\! \! \perp} \right) + \partial_{\! \! \perp} \otimes \rnu \otimes \rnu \otimes \partial_{\! \! \perp} \right\}
\end{eqnarray}

\noindent together with the $\Re$-term (\ref{Re(n-lambda-n)}), we divide contributions to $\Delta {\cal M}$ into three types. The first term in the braces in (\ref{Im(nn)Im(nn)}) corresponds to the contribution of the type $\rnu \dots \rnu$ in which each $\rnu$ is contracted with the corresponding $w$ from $w \Delta {\cal M} w$. The second term in the braces in (\ref{Im(nn)Im(nn)}) corresponds to the contribution of the type $\partial_{\! \! \perp} \dots \rnu + \rnu \dots \partial_{\! \! \perp}$ in which one $\rnu$ should be contracted with one of the two $w$s from $w \Delta {\cal M} w$ and one $\partial_{\! \! \perp}$ should be contracted with the other $w$. Finally, the third term in the braces in (\ref{Im(nn)Im(nn)}) corresponds to the contribution of the type $\partial_{\! \! \perp} \dots \partial_{\! \! \perp}$ in which each $\partial_{\! \! \perp}$ is contracted with the corresponding $w$ from $w \Delta {\cal M} w$.

First consider the contribution of the type $\rnu \dots \rnu$. There may also be two $\partial_{\! \! \perp}$s which are contracted with $G^{\rm aux}$ and hence with $G^{(0)}$. There each $G^{(0)}$ enters, being contracted with $\partial_{\! \! \perp}$ on both sides, as $\partial_{\! \! \perp}^\mu G^{(0)}_{\lambda \mu \sigma \tau} \partial_{\! \! \perp}^\sigma$ and can thus be written in terms of $M$ by the definition of $M$ (\ref{a-dGd=M}). We also take into account the $\rnu \dots \rnu$ part of the $\Re$-term in $\Delta {\cal M}$ and obtain
\begin{eqnarray}\label{nu...nu}                                            %25
& & \Delta {\cal M }^{\lambda \mu \sigma \tau}_{\rnu \dots \rnu} = - \frac{1}{2} \rnu^\mu \rlambda^{\! \! \lambda \sigma} \rnu^\tau - \frac{ \varepsilon^2 }{4} \rnu^\mu \rlambda^{\! \! \lambda \zeta} \frac{ \partial_{\! \! \perp}^\pi }{ \partial_{\! \! \perp}^2 } \left\{ G^{(0)}_{\zeta \pi \chi \psi} \right. \nonumber \\ & & \left. + \frac{(\rnu \partial )^2}{2} G^{(0)}_{\zeta \pi \nu \rho } \frac{ \partial_{\! \! \perp}^\nu }{ \partial_{\! \! \perp}^2 } ( M^{- 1} )^{ \rho \kappa } \frac{ \partial_{\! \! \perp}^\varphi }{ \partial_{\! \! \perp}^2 } G^{(0)}_{\kappa \varphi \chi \psi} \right\} \frac{ \partial_{\! \! \perp}^\chi }{ \partial_{\! \! \perp}^2 } \rlambda^{\! \! \psi \sigma} \rnu^\tau = - \frac{1}{2} \rnu^\mu \rlambda^{\! \! \lambda \sigma} \rnu^\tau \nonumber \\ & &  - \frac{ \varepsilon^2 }{4} \rnu^\mu \rlambda^{\! \! \lambda \zeta} \left[ \frac{2}{ \varepsilon^2 } \ralpha_{ \zeta \psi } - \frac{2}{(\rnu \partial )^2} M_{ \zeta \psi } \right] \rlambda^{\! \! \psi \sigma} \rnu^\tau - \frac{ \varepsilon^2 }{4} \rnu^\mu \rlambda^{\! \! \lambda \zeta} \frac{(\rnu \partial )^2}{2} \left[ \frac{2}{ \varepsilon^2 } \ralpha_{ \zeta \rho } - \frac{2}{(\rnu \partial )^2} M_{ \zeta \rho } \right] \nonumber \\ & &  \cdot ( M^{- 1} )^{ \rho \kappa } \left[ \frac{2}{ \varepsilon^2 } \ralpha_{ \kappa \psi } - \frac{2}{(\rnu \partial )^2} M_{ \kappa \psi } \right] \rlambda^{\! \! \psi \sigma} \rnu^\tau = - \frac{(\rnu \partial )^2}{2 \varepsilon^2 } \rnu^\mu ( M^{- 1} )^{ \lambda \sigma } \rnu^\tau .
\end{eqnarray}

\noindent Symmetrization in pairs $( \lambda \mu )$ and $( \sigma \tau )$ is implied. However, this is not necessary, since further such expressions will already be contracted here with the symmetric tensors $g_{ \lambda \mu }$ and $g_{ \sigma \tau }$.

Now let's pass to the $\partial_{\! \! \perp} \dots \rnu + \rnu \dots \partial_{\! \! \perp}$ part of $\Delta {\cal M}$. Here a simplification occurs when contracting the propagator $G^{(0)}$ with $\rnu$,
\begin{equation}                                                           %26
\rnu^\mu G^{(0)}_{\lambda \mu \sigma \tau} = - \frac{2}{(\rnu \partial )^2} \left[ (\rnu \partial ) (\ralpha_{\lambda \sigma} \partial_\tau + \ralpha_{\lambda \tau} \partial_\sigma ) - \ralpha_{\lambda \mu} \rnu^\mu \partial_\sigma \partial_\tau \right],
\end{equation}

\noindent since then contracting with $\rlambda^{\! \rho \lambda}$ over the index $\lambda$ completely cancels the dependence on $\| \ralpha \|$,
\begin{equation}                                                           %27
\rlambda^{\! \rho \lambda} \rnu^\mu G^{(0)}_{\lambda \mu \sigma \tau} = - \frac{2}{(\rnu \partial )^2} \left[ (\rnu \partial ) (\delta^\rho_\sigma \partial_\tau + \delta^\rho_\tau \partial_\sigma ) - \rnu^\rho \partial_\sigma \partial_\tau \right].
\end{equation}

\noindent Then contracting with $\partial_{\! \! \perp}^\sigma$ gives
\begin{equation}                                                           %28
\rlambda^{\! \rho \lambda} \rnu^\mu G^{(0)}_{\lambda \mu \sigma \tau} \partial_{\! \! \perp}^\sigma = - \frac{2 \partial_{\! \! \perp}^2 }{\rnu \partial } \left[ \delta^\rho_\tau + \left( \frac{ \partial_{\! \! \perp}^\rho }{ \partial_{\! \! \perp}^2 } - \frac{ \rnu^\rho }{ \rnu \partial } \right) \partial_\tau \right].
\end{equation}

\noindent When finding $\partial_{\! \! \perp} \dots \rnu$ part of $\Delta {\cal M}$ we encounter values of this type; also expressing the value $\partial_{\! \! \perp}^\mu G^{(0)}_{\lambda \mu \sigma \tau} \partial_{\! \! \perp}^\sigma$ in terms of $M$ as above, we find
\begin{eqnarray}                                                           %29
& & \Delta {\cal M }^{\lambda \mu \sigma \tau}_{\partial_{\! \! \perp} \dots \rnu} = - \frac{ \varepsilon^2 }{4} \frac{ \partial_{\! \! \perp}^\mu }{ \partial_{\! \! \perp}^2 } \rlambda^{\! \! \lambda \zeta} \rnu^\pi \left\{ G^{(0)}_{\zeta \pi \chi \psi} + \frac{(\rnu \partial )^2}{2} G^{(0)}_{\zeta \pi \nu \rho } \frac{ \partial_{\! \! \perp}^\nu }{ \partial_{\! \! \perp}^2 } ( M^{- 1} )^{ \rho \kappa } \frac{ \partial_{\! \! \perp}^\varphi }{ \partial_{\! \! \perp}^2 } G^{(0)}_{\kappa \varphi \chi \psi} \right\} \frac{ \partial_{\! \! \perp}^\chi }{ \partial_{\! \! \perp}^2 } \nonumber \\ & & \cdot \rlambda^{\! \! \psi \sigma} \rnu^\tau = - \frac{ \varepsilon^2 }{4} \frac{ \partial_{\! \! \perp}^\mu }{ \partial_{\! \! \perp}^2 } \rlambda^{\! \! \lambda \zeta} \rnu^\pi \left\{ G^{(0)}_{\zeta \pi \chi \psi} \frac{ \partial_{\! \! \perp}^\chi }{ \partial_{\! \! \perp}^2 } + (\rnu \partial )^2 G^{(0)}_{\zeta \pi \nu \rho } \frac{ \partial_{\! \! \perp}^\nu }{ \partial_{\! \! \perp}^2 } ( M^{- 1} )^{ \rho \kappa } \left[ \frac{1}{ \varepsilon^2 } \ralpha_{ \kappa \psi } \right. \right. \nonumber \\ & & \left. \left. - \frac{1}{(\rnu \partial )^2} M_{ \kappa \psi } \right] \right\} \rlambda^{\! \! \psi \sigma} \rnu^\tau = - \frac{(\rnu \partial )^2}{4} \frac{ \partial_{\! \! \perp}^\mu }{ \partial_{\! \! \perp}^2 } \rlambda^{\! \! \lambda \zeta} \rnu^\pi G^{(0)}_{\zeta \pi \nu \rho } \frac{ \partial_{\! \! \perp}^\nu }{ \partial_{\! \! \perp}^2 } ( M^{- 1} )^{ \rho \sigma } \rnu^\tau.
\end{eqnarray}

\noindent Together with the part $\rnu \dots \partial_{\! \! \perp}$ that looks symmetrical relative to $\partial_{\! \! \perp} \dots \rnu$, this is written as follows:
\begin{eqnarray}\label{d...nu+nu...d}                                      %30
\Delta {\cal M }^{\lambda \mu \sigma \tau}_{\partial_{\! \! \perp} \dots \rnu + \rnu \dots \partial_{\! \! \perp}} & = & - \frac{(\rnu \partial )^2}{4} \frac{ \partial_{\! \! \perp}^\mu }{ \partial_{\! \! \perp}^2 } \rlambda^{\! \! \lambda \zeta} \rnu^\pi G^{(0)}_{\zeta \pi \nu \rho } \frac{ \partial_{\! \! \perp}^\nu }{ \partial_{\! \! \perp}^2 } ( M^{- 1} )^{ \rho \sigma } \rnu^\tau \nonumber \\ & & - \frac{(\rnu \partial )^2}{4} \rnu^\mu ( M^{- 1} )^{ \lambda \kappa } \frac{ \partial_{\! \! \perp}^\varphi }{ \partial_{\! \! \perp}^2 } G^{(0)}_{\kappa \varphi \chi \psi} \rnu^\chi \rlambda^{\! \! \psi \sigma} \frac{ \partial_{\! \! \perp}^\tau }{ \partial_{\! \! \perp}^2 } .
\end{eqnarray}

Finally, in the $\partial_{\! \! \perp} \dots \partial_{\! \! \perp}$ part of $\Delta {\cal M}$,
\begin{eqnarray}                                                           %31
\Delta {\cal M }^{\lambda \mu \sigma \tau}_{\partial_{\! \! \perp} \dots \partial_{\! \! \perp}} & = & - \frac{ \varepsilon^2 }{2} \frac{ \partial_{\! \! \perp}^\mu }{ \partial_{\! \! \perp}^2 } \rlambda^{\! \! \lambda \sigma} \frac{ \partial_{\! \! \perp}^\tau }{ \partial_{\! \! \perp}^2 } - \frac{ \varepsilon^2 }{4} \frac{ \partial_{\! \! \perp}^\mu }{ \partial_{\! \! \perp}^2 } \rlambda^{\! \! \lambda \zeta} \rnu^\pi \left\{ G^{(0)}_{\zeta \pi \chi \psi} \right. \nonumber \\ & & \left. + \frac{(\rnu \partial )^2}{2} G^{(0)}_{\zeta \pi \nu \rho } \frac{ \partial_{\! \! \perp}^\nu }{ \partial_{\! \! \perp}^2 } ( M^{- 1} )^{ \rho \kappa } \frac{ \partial_{\! \! \perp}^\varphi }{ \partial_{\! \! \perp}^2 } G^{(0)}_{\kappa \varphi \chi \psi} \right\} \rnu^\chi \rlambda^{\! \! \psi \sigma} \frac{ \partial_{\! \! \perp}^\tau }{ \partial_{\! \! \perp}^2 },
\end{eqnarray}

\noindent it is important that the first term (the $\varepsilon^2$-part of the $\Re ( n^\mu \rlambda^{\! \! \lambda \sigma } n^\tau )$-term) and the first term in the braces (linear in $G^{(0)}_{\zeta \pi \chi \psi}$) prove to mutually cancel each other, so that we have an expression of the type $ \dots M^{- 1} \dots $, as for the other parts of $\Delta {\cal M}$:
\begin{equation}\label{d...d}                                              %32
\Delta {\cal M }^{\lambda \mu \sigma \tau}_{\partial_{\! \! \perp} \dots \partial_{\! \! \perp}} = - \frac{ \varepsilon^2 }{8} (\rnu \partial )^2 \frac{ \partial_{\! \! \perp}^\mu }{ \partial_{\! \! \perp}^2 } \rlambda^{\! \! \lambda \zeta} \rnu^\pi G^{(0)}_{\zeta \pi \nu \rho } \frac{ \partial_{\! \! \perp}^\nu }{ \partial_{\! \! \perp}^2 } ( M^{- 1} )^{ \rho \kappa } \frac{ \partial_{\! \! \perp}^\varphi }{ \partial_{\! \! \perp}^2 } G^{(0)}_{\kappa \varphi \chi \psi} \rnu^\chi \rlambda^{\! \! \psi \sigma} \frac{ \partial_{\! \! \perp}^\tau }{ \partial_{\! \! \perp}^2 }.
\end{equation}

Collecting the obtained parts (\ref{nu...nu}), (\ref{d...nu+nu...d}) and (\ref{d...d}), we find the additional term to the matrix of the bilinear form in the gravity action,
\begin{eqnarray}                                                           %33
\Delta {\cal M }^{\lambda \mu \sigma \tau} & = & - \frac{ (\rnu \partial )^2 }{ 2 \varepsilon^2 } \left( \rnu^\mu \delta^\lambda_\rho + \frac{ \varepsilon^2 }{ 2 } \frac{ \partial_{\! \! \perp}^\mu }{ \partial_{\! \! \perp}^2 } \rlambda^{\! \! \lambda \zeta} \rnu^\pi G^{(0)}_{\zeta \pi \nu \rho } \frac{ \partial_{\! \! \perp}^\nu }{ \partial_{\! \! \perp}^2 } \right) ( M^{- 1} )^{ \rho \kappa } \nonumber \\ & & \cdot \left( \delta^\sigma_\kappa \rnu^\tau + \frac{ \varepsilon^2 }{ 2 } \frac{ \partial_{\! \! \perp}^\varphi }{ \partial_{\! \! \perp}^2 } G^{(0)}_{\kappa \varphi \chi \psi} \rnu^\chi \rlambda^{\! \! \psi \sigma} \frac{ \partial_{\! \! \perp}^\tau }{ \partial_{\! \! \perp}^2 } \right) .
\end{eqnarray}

\section{Ghost contribution}\label{ghost}

Thus, the gauge non-invariant term corresponding to the principal value synchronous propagator is
\begin{eqnarray}\label{gauge-viol}                                         %34
\Delta S = - \frac{1}{4} \int \d^4 x f_\rho [ g ] \Lambda^{\rho \kappa} f_\kappa [ g ] , ~~~ \Lambda = \frac{ ( \rnu \partial )^2 }{ \varepsilon^2 } M^{- 1} , ~~~ f_\rho = {\rm O }^{\lambda \mu}_\rho w_{\lambda \mu} ,\nonumber \\ {\rm O }^{\lambda \mu}_\rho = \delta^\lambda_\rho \rnu^\mu + \frac{ \varepsilon^2 }{ 2 } \frac{ \partial_{\! \! \perp}^\nu }{ \partial_{\! \! \perp}^2 } G^{(0)}_{\rho \nu \pi \zeta} \rnu^\pi \rlambda^{\! \! \zeta \lambda} \frac{ \partial_{\! \! \perp}^\mu }{ \partial_{\! \! \perp}^2 } = \delta^\lambda_\rho \rnu^\mu - \frac{ \varepsilon^2 }{ \rnu \partial } \left[ \delta^\lambda_\rho + \partial_\rho \left( \frac{ \partial_{\! \! \perp}^\lambda }{ \partial_{\! \! \perp}^2 } - \frac{ \rnu^\lambda }{ \rnu \partial } \right) \right] \frac{ \partial_{\! \! \perp}^\mu }{ \partial_{\! \! \perp}^2 }.
\end{eqnarray}

Analogously to a gauge theory with the gauge violating term with $\Lambda^{\rho \kappa} \sim \eta^{\rho \kappa}$, we consider a general gauge condition of the form
\begin{equation}                                                           %35
f_\lambda [ g ] ( x ) = \rkappa_\lambda ( x )
\end{equation}

\noindent characterized by a vector function $\rkappa_\lambda ( x )$. The functional integral acquires the delta-function factor $\prod_{x , \lambda} \delta (f^\lambda [ g ] ( x ) - \rkappa^\lambda ( x ) )$ together with the corresponding normalization factor $\Phi_0 [ g ]$. On the one hand, the functional integral does not depend on the gauge defined here by $\rkappa_\lambda ( x )$, on the other hand, it can be averaged by means of the functional integration over $\prod_{x , \lambda} \d \rkappa_\lambda ( x )$ with the weight factor of the form $\exp \left ( - \frac{ i }{ 4 } \int \rkappa_\rho ( x ) \Lambda^{\rho \kappa} \rkappa_\kappa ( x ) \d^4 x \right )$, thereby reproducing the gauge violating term $ - \frac{ 1 }{ 4 } \int f_\rho [ g ] ( x ) \Lambda^{\rho \kappa} f_\kappa [ g ] ( x ) \d^4 x $ in the action. As for the factor $\Phi_0 [ g ]$, it is equal to $\Det {\cal O}$, where the operator ${\cal O}$ is defined by subjecting $f_\lambda [ g ]$ to the infinitesimal gauge (diffeomorphism) transformation
\begin{equation}                                                           %36
\Xi : ~~~ \delta x^\lambda = \xi^\lambda ( x ) , ~~~ \delta g_{\lambda \mu} = (g_{\lambda \mu})^\Xi - g_{\lambda \mu} = - \xi_{\lambda , \mu} - \xi_{\mu , \lambda} + 2 \Gamma^\nu_{\lambda \mu} \xi_\nu ,
\end{equation}

\noindent so that
\begin{equation}                                                           %37
\delta f_\rho = {\rm O }^{\lambda \mu}_\rho \delta g_{\lambda \mu} \equiv - {\cal O}_\rho{}^\nu \xi_\nu .
\end{equation}

\noindent Thus,
\begin{eqnarray}\label{calO}                                               %38
{\cal O}_\rho{}^\nu = \left( \rnu \partial - \frac{ \varepsilon^2 }{ \rnu \partial } \right) \delta_\rho^\nu + \left[ 1 + \frac{ \varepsilon^2 }{ (\rnu \partial)^2 } \right] \partial_\rho \rnu^\nu - \frac{ 2 \varepsilon^2 }{ (\rnu \partial) \partial_{\! \! \perp}^2 } \partial_\rho \partial_{\! \! \perp}^\nu \nonumber \\ - 2 \rnu^\mu \Gamma^\nu_{\rho \mu} + \frac{ 2 \varepsilon^2 }{ (\rnu \partial) \partial_{\! \! \perp}^2 } \partial_{\! \! \perp}^\mu \Gamma^\nu_{\rho \mu} + \frac{ 2 \varepsilon^2 }{ (\rnu \partial) \partial_{\! \! \perp}^2 } \partial_\rho \left( \frac{ \partial_{\! \! \perp}^\lambda }{ \partial_{\! \! \perp}^2 } - \frac{ \rnu^\lambda }{ \rnu \partial } \right) \partial_{\! \! \perp}^\mu \Gamma^\nu_{\lambda \mu}.
\end{eqnarray}

Denoting the free part of (\ref{calO}) (i. e., when $\Gamma^\nu_{\lambda \mu} = 0$, top line) as ${\cal O}_{ ( 0 ) \rho}{}^\nu$,
\begin{equation}                                                           %39
{\rm O }^{\lambda \mu}_\rho ( \xi_{\lambda , \mu} + \xi_{\mu , \lambda} ) \equiv {\cal O}_{ ( 0 ) \rho}{}^\nu \xi_\nu ,
\end{equation}

\noindent we find
\begin{equation}                                                           %40
{\cal O}_{ ( 0 ) \nu}^{ - 1 }{}^\rho = \frac{ \rnu \partial }{(\rnu \partial)^2 - \varepsilon^2} \delta_\rnu{}^\rho - \frac{1}{2} \frac{ (\rnu \partial)^2 + \varepsilon^2 }{[ (\rnu \partial)^2 - \varepsilon^2 ]^2} \partial_\nu \rnu^\rho + \frac{ \varepsilon^2 \rnu \partial }{[ (\rnu \partial)^2 - \varepsilon^2 ]^2} \frac{ \partial_\nu \partial_{\! \! \perp}^\rho }{ \partial_{\! \! \perp}^2 }.
\end{equation}

\noindent (If we introduce the ghost Lagrangian density $\otheta^\mu {\cal O}_\mu{}^\lambda \vartheta_\lambda$ with anticommuting vector fields $\vartheta_\lambda$, $\otheta^\lambda$, then ${\cal O}_{ ( 0 ) }^{ - 1 }$ will be a ghost propagator.) The perturbative expansion for $\Det {\cal O}$ follows when factoring out the normalization constant $\Det {\cal O}_{ ( 0 ) }$ and expanding $\ln \Det ( {\cal O}_{ ( 0 ) }^{ - 1 } {\cal O} )$, where
\begin{eqnarray}\label{O/O(0)}                                             %41
& & \hspace{-7mm} ( {\cal O}_{ ( 0 ) }^{ - 1 } {\cal O} )_\rho{}^\nu = \delta_\rho{}^\nu - \frac{2}{(\rnu \partial)^2 - \varepsilon^2} \left[ (\rnu \partial) \rnu^\mu - \varepsilon^2 \frac{ \partial_{\! \! \perp}^\mu }{ \partial_{\! \! \perp}^2 } \right] \Gamma^\nu_{\rho \mu} \nonumber \\ & & \hspace{-7mm} + \left\{ \frac{ (\rnu \partial)^2 + \varepsilon^2 }{[ (\rnu \partial)^2 - \varepsilon^2 ]^2} \left[ \rnu^\lambda \rnu^\mu + \varepsilon^2 \frac{ \partial_{\! \! \perp}^\lambda \partial_{\! \! \perp}^\mu }{ ( \partial_{\! \! \perp}^2 )^2 } \right] - \frac{ 4 \varepsilon^2 ( \rnu \partial ) \rnu^\lambda }{[ (\rnu \partial)^2 - \varepsilon^2 ]^2} \frac{ \partial_{\! \! \perp}^\mu }{ \partial_{\! \! \perp}^2 } \right\} \partial_\rho \Gamma^\nu_{\lambda \mu} \equiv \delta_\rho{}^\nu + W_\rho{}^\nu.
\end{eqnarray}

\noindent Then
\begin{equation}                                                           %42
\ln \Det ( {\cal O}_{ ( 0 ) }^{ - 1 } {\cal O} ) = \Tr \ln ( 1 + W ) = \sum_n \frac{ ( -1 )^{n + 1}}{n} \Tr ( W^n ).
\end{equation}

\noindent Here the $n$-th term corresponds to ghost diagrams with $n$ external lines.

In addition to the explicit smallness $\varepsilon^2$, here we have a smallness indirectly caused by the product of $\rnu$ and the metric of Christoffel symbols playing the role of external fields. This smallness is provided by the factor $\exp \left ( - \frac{ i }{ 4 } \int f_\rho [ g ] ( x ) \Lambda^{\rho \kappa} f_\kappa [ g ] ( x ) \d^4 x \right )$ in the functional integral with $\Lambda = ( \rnu \partial )^2 M^{ - 1 } \varepsilon^{ - 2 }$, so that a typical $\rnu^\mu g_{\lambda \mu} = O ( \sqrt{ \varepsilon^2 } ) = O ( | \varepsilon | ) = O ( \varepsilon )$. These $\rnu^\mu g_{\lambda \mu}$ enter the ghost action linearly through $ \rnu^\mu \Gamma^\nu_{\rho \mu} $ and $\rnu^\lambda \rnu^\mu \Gamma^\nu_{\lambda \mu} $ and, generally speaking, may result in terms $O( \varepsilon )$ in the effective ghost action $S_{\rm ghost} = - i \ln \Det {\cal O}$. Here it is also important that $\Det M$ be finite at a sufficient number of points almost everywhere at $\varepsilon \to 0$.

So we have a small value
\begin{equation}\label{O(e)}                                               %43
\rnu^\lambda \rnu^\mu \Gamma^\nu_{\lambda \mu} = \rnu^\lambda \rnu^\mu g^{\nu \sigma} g_{\sigma \lambda, \mu} + O( \varepsilon^2 ) = O( \varepsilon )
\end{equation}

\noindent (it is taken into account that $\rnu^\lambda \rnu^\mu g_{\lambda \mu, \sigma } = O( \varepsilon^2 )$). Besides that, the external field $ \rnu^\mu \Gamma^\nu_{\rho \mu} $, which has order $O(1)$, can be expanded according to the formula
\begin{equation}\label{O(1)+O(e)}                                          %44
2 \rnu^\mu \Gamma^\nu_{\rho \mu} = g^{\nu \lambda} g_{\lambda \rho, \mu} \rnu^\mu + g^{\nu \lambda} ( g_{\lambda \mu, \rho} - g_{\rho \mu, \lambda} ) \rnu^\mu,
\end{equation}

\noindent where the first and second terms have order $O(1)$ and $O( \varepsilon )$, respectively.

The largest $O(1)$ contribution is a priori expected from diagrams with an arbitrary number of external lines due solely to the field of order $O(1)$ in (\ref{O(1)+O(e)}); the $O( \varepsilon )$ contribution arises if one of these lines is due to an external field of order $O( \varepsilon )$. Let us first consider the case when there are $(n - 1)$ external lines of the field $g^{\nu \lambda} g_{\lambda \rho, \mu} \rnu^\mu$ and one external line of the field $\rnu^\lambda \rnu^\mu g^{\nu \sigma} g_{\sigma \lambda, \mu} = O( \varepsilon )$ (\ref{O(e)}). These fields appear in the expression for the diagram as
\begin{eqnarray}                                                           %45
g^{\nu_1 \lambda_2} g_{\lambda_2 \nu_2, \mu_2} \rnu^{\mu_2} \otimes g^{\nu_2 \lambda_3} g_{\lambda_3 \nu_3, \mu_3} \rnu^{\mu_3} \otimes \dots \otimes g^{\nu_{(j - 1)} \lambda_j} g_{\lambda_j \nu_j, \mu_j} \rnu^{\mu_j} \otimes \dots \nonumber \\ \dots \otimes g^{\nu_{(n - 1)} \lambda_n} g_{\lambda_n \nu_n, \mu_n} \rnu^{\mu_n} \otimes \partial_{\nu_1} g^{\nu_n \sigma} g_{\sigma \lambda_1, \mu_1} \rnu^{\lambda_1} \rnu^{\mu_1} \nonumber \\ = \left[ \bigotimes_{j = 2}^n g^{\nu_{(j - 1)} \lambda_j} g_{\lambda_j \nu_j, \mu_j} \rnu^{\mu_j} \right]_{\nu_n}^{~~\nu_1} \otimes \left[ \partial_{\! \! \perp \nu_1} + \frac{ \rnu \partial }{\rnu^2 } \rnu_{\nu_1} \right] g^{\nu_n \sigma} g_{\sigma \lambda_1, \mu_1} \rnu^{\lambda_1} \rnu^{\mu_1} .
\end{eqnarray}

\noindent Here the factors in the tensor product are generally taken at different coordinates (here $x^0$), but the factors of the type $g^{\nu \lambda} g_{\lambda \rho, \mu} \rnu^\mu$ have a common feature: they are a matrix with elements $O( \varepsilon )$ at $\nu = 0$ or $\rho = 0$ ($O( \varepsilon^2 )$ at $\nu = \rho = 0$) or $O(1)$ at $\nu \neq 0$, $\rho \neq 0$. Taking the product of such matrices in the tensor product $\bigotimes_{j = 2}^n$ gives a matrix with indices $\nu_1$, $\nu_n$ with the same property, in particular, it is $O(1)$ only for $\nu_1 \neq 0$ (and $\nu_n \neq 0$). For $\nu_1 \neq 0$ we have $\partial_{\nu_1} = \partial_{\! \! \perp \nu_1}$, and since in the considered diagram there is no additional dependence on $p_{\! \! \perp \nu_1} = - i \partial_{\! \! \perp \nu_1}$, then integration over $\d^3 p_{\! \! \perp \nu_1}$ gives zero. Thus, the overall diagram is $O( \varepsilon^2 )$.

Next, we will take into account an arbitrary number of external lines of the field $g^{\nu \lambda} g_{\lambda \rho, \mu} \rnu^\mu$ and consider what the order of magnitude of the contribution of the field $g^{\nu \lambda} ( g_{\lambda \mu, \rho} - g_{\rho \mu, \lambda} ) \rnu^\mu$ will be. To do this, we write the $O(1)$- (${\bf 1}$ and $g^{\nu \lambda} g_{\lambda \rho, \mu} \rnu^\mu$) and $g^{\nu \lambda} ( g_{\lambda \mu, \rho} - g_{\rho \mu, \lambda} ) \rnu^\mu$-parts of the operator $( {\cal O}_{ ( 0 ) }^{ - 1 } {\cal O} )_\rho{}^\nu$ (\ref{O/O(0)}), multiplied on the right by the metric matrix $\| g \|$, as the sum of the symmetric $S_{\rho \nu}$ (from ${\bf 1}$ and $g^{\nu \lambda} g_{\lambda \rho, \mu} \rnu^\mu$)  and antisymmetric $A_{\rho \nu}$ (from $g^{\nu \lambda} ( g_{\lambda \mu, \rho} - g_{\rho \mu, \lambda} ) \rnu^\mu$) parts,
\begin{eqnarray}                                                           %46
& & ( {\cal O}_{ ( 0 ) }^{ - 1 } {\cal O} \| g \| )_{\rho \nu} = S_{\rho \nu} + A_{\rho \nu} + \dots , ~~~ S_{\rho \nu} = g_{\rho \nu} - \frac{ ( \rnu \partial ) \rnu^\mu }{( \rnu \partial )^2 - \varepsilon^2 } g_{\rho \nu, \mu}, \nonumber \\ & & A_{\rho \nu} = \frac{ ( \rnu \partial ) \rnu^\mu }{( \rnu \partial )^2 - \varepsilon^2 } ( g_{\rho \mu, \nu} - g_{\nu \mu, \rho} ).
\end{eqnarray}

\noindent Given $\Det \| g \| = \prod_x \det \| g \|$, this yields $\Det ( {\cal O}_{ ( 0 ) }^{ - 1 } {\cal O} )$ according to
\begin{eqnarray}                                                           %47
\ln \left[ \Det \| g \| \Det ( {\cal O}_{ ( 0 ) }^{ - 1 } {\cal O} ) \right] = \Tr \ln ( S + A + \dots ) \nonumber \\ = \Tr \ln S + \Tr \left( S^{- 1} A \right) - \frac{1}{2} \Tr \left( S^{- 1} A S^{- 1} A \right) + \dots .
\end{eqnarray}

\noindent Here $\Tr \left( S^{- 1} A \right) = 0$, and therefore the contribution of the diagrams for $\ln \Det ( {\cal O}_{ ( 0 ) }^{ - 1 } {\cal O} )$ with external field lines $g^{\nu \lambda} ( g_{\lambda \mu, \rho} - g_{\rho \mu, \lambda} ) \rnu^\mu$ starts from the order $O( \varepsilon^2 )$.

Finally, the contribution of diagrams with an arbitrary number of external lines, due solely to the field $g^{\nu \lambda} g_{\lambda \rho, \mu} \rnu^\mu$, follows, if we restrict ourselves to the part of ${\cal O}_{ ( 0 ) }^{ - 1 } {\cal O}$,
\begin{eqnarray}                                                           %48
( {\cal O}_{ ( 0 ) }^{ - 1 } {\cal O} )_\rho{}^\nu = S_{\rho \lambda} g^{\lambda \nu} + \dots = \delta_\rho{}^\nu - \frac{ ( \rnu \partial ) \rnu^\mu }{( \rnu \partial )^2 - \varepsilon^2 } g_{\rho \lambda, \mu} g^{\lambda \nu} + \dots \nonumber \\ = \frac{1}{( \rnu \partial )^2 - \varepsilon^2 } \left[ ( \rnu \partial ) g_{\rho \lambda} ( \rnu \partial ) g^{\lambda \nu} - \varepsilon^2 \delta_\rho{}^\nu \right] + \dots ,
\end{eqnarray}

\noindent so that
\begin{equation}                                                           %49
\ln \Det ( {\cal O}_{ ( 0 ) }^{ - 1 } {\cal O} ) = \ln \frac{ \Det \left[ ( \rnu \partial ) \| g \| ( \rnu \partial ) \| g \|^{- 1} - \varepsilon^2 \right] }{ \Det \left[ ( \rnu \partial )^2 - \varepsilon^2 \right] } + \dots .
\end{equation}

\noindent Roughly, if we tend $\varepsilon \to 0$, then we arrive at the determinant of the product of operators $( \rnu \partial ) \| g \| ( \rnu \partial ) \| g \|^{- 1}$, which is factorized, and any dependence on $\| g \|$ disappears. More strictly, we can adopt a detailed expression for unbounded operators in the implied discrete model for the functional integral. Whereas the above consideration was aimed at eliminating IR divergences, then now, reducing the matter to the calculation of $\Det \left[ ( \rnu \partial ) \| g \| ( \rnu \partial ) \| g \|^{- 1} - \varepsilon^2 \right]$, the important thing is to resolve the behavior at small distances, namely, the finite-difference form of the derivative. It is natural to adopt that $\partial$ is a triangular matrix over points (say, a lower triangular matrix); $\| g \|$ is a diagonal matrix over points $x$ (more exactly, block-diagonal over pairs of points $x$ and indices $\lambda$). The product of four such matrices is again a triangular matrix, and we have the determinant of a triangular matrix. Such a determinant is a product of the diagonal elements of the matrix, in each of these elements we have the product of $\| g \|$ and $\| g \|^{- 1}$ at the same point, that is ${\bf 1}$. Thus, the dependence on $\| g \|$ disappears (here also for $\varepsilon \neq 0$), and we have zero for the contribution to $\ln \Det ( {\cal O}_{ ( 0 ) }^{ - 1 } {\cal O} )$ from the diagrams containing only external field lines $g^{\nu \lambda} g_{\lambda \rho, \mu} \rnu^\mu$.

Thus, the effective ghost action is of order $O( \varepsilon^2 )$ at $\varepsilon \to 0$.

If we actually compute the diagrams for the effective ghost action, we will encounter divergent diagrams (UV-divergent when integrated over $\d^3 p_{\! \! \perp} = \d^3 \bp$) that are regularized in the underlying discrete theory. Further, the result of such a calculation is a functional of the external field $\Gamma^\lambda_{\mu \nu}$. An upper bound for this field and for the effective ghost action itself follows, as we consider in \cite{khat1}, if we turn again to the discrete theory and to regularizations of the type of spatial and temporal coordinate lattice steps, maximum spatial and temporal lengths, a minimum 3-volume, a large but finite number of points appearing in the phenomenological definition of the functional integral. Then we tend $\varepsilon$ to 0.

The dependence on $\rlambda^{\lambda \mu}$ (or $\ralpha_{\lambda \mu}$) is concentrated in the matrix $M$. Consider its typical form. It is natural to consider the form of $\ralpha_{\lambda \mu}$, corresponding to the adopted symmetry and consisting of a few structures that can be constructed from $\eta_{\lambda \mu}$ and the two 4-vectors $\partial_\lambda$, $\rnu_\lambda$ or $\partial_{\! \! \perp \lambda}$, $\rnu_\lambda$ at our disposal,
\begin{eqnarray}                                                           %50
& & \ralpha_{\lambda \mu} = \ralpha_P P_{\lambda \mu} + \ralpha_\rnu \frac{ \rnu_\lambda \rnu_\mu }{ \rnu^2 } + \ralpha_\partial \frac{ \partial_{\! \! \perp \lambda} \partial_{\! \! \perp \mu } }{ \partial^2_{\! \! \perp } } + \ralpha_{\rnu \partial + } ( \rnu \partial ) ( \rnu_\lambda \partial_{\! \! \perp \mu } + \partial_{\! \! \perp \lambda } \rnu_\mu) \nonumber \\ & & + \ralpha_{\rnu \partial - } ( \rnu_\lambda \partial_{\! \! \perp \mu } - \partial_{\! \! \perp \lambda } \rnu_\mu), \nonumber \\ & & P_{\lambda \mu} = \eta_{\lambda \mu} - \frac{ \rnu_\lambda \rnu_\mu }{ \rnu^2 } - \frac{ \partial_{\! \! \perp \lambda} \partial_{\! \! \perp \mu } }{ \partial^2_{\! \! \perp } }.
\end{eqnarray}

\noindent Here $\ralpha_P$, $\ralpha_\rnu$, $\ralpha_\partial$, $\ralpha_{\rnu \partial + }$, $\ralpha_{\rnu \partial - }$ can be Hermitian symmetric operators (functions of $\partial$).

\noindent The matrix $M$ reads
\begin{eqnarray}                                                           %51
& & M_{\lambda \mu} = \rmu_P P_{\lambda \mu} + \rmu_\rnu \frac{ \rnu_\lambda \rnu_\mu }{ \rnu^2 } + \rmu_\partial \frac{ \partial_{\! \! \perp \lambda} \partial_{\! \! \perp \mu } }{ \partial^2_{\! \! \perp } } + \rmu_{\rnu \partial + } ( \rnu \partial ) ( \rnu_\lambda \partial_{\! \! \perp \mu } + \partial_{\! \! \perp \lambda } \rnu_\mu) \nonumber \\ & & + \rmu_{\rnu \partial - } ( \rnu_\lambda \partial_{\! \! \perp \mu } - \partial_{\! \! \perp \lambda } \rnu_\mu), \nonumber \\ & & \rmu_P = \ralpha_P \frac{ ( \rnu \partial )^2 }{ \varepsilon^2 } - \frac{ \rnu^2 }{ \partial_{\! \! \perp}^2} + \ralpha_P, ~ \rmu_\rnu = \ralpha_\rnu \frac{ ( \rnu \partial )^2 }{ \varepsilon^2 } + \ralpha_\partial \frac{ ( \rnu \partial )^2 }{ \rnu^2 \partial_{\! \! \perp}^2 }, \nonumber \\ & & \rmu_\partial = - \frac{ \rnu^4 }{ ( \rnu \partial )^2 } \frac{ \partial^2 }{ \partial_{\! \! \perp}^2 } + \ralpha_\partial \frac{ ( \rnu \partial )^2 }{ \varepsilon^2 } + \ralpha_\rnu \frac{ \rnu^2 \partial_{\! \! \perp}^2 }{ ( \rnu \partial )^2 } + 4 \left( \ralpha_\partial - \ralpha_{\rnu \partial +} \rnu^2 \partial_{\! \! \perp}^2 \right), \nonumber \\ & & \rmu_{\rnu \partial +} = \ralpha_{\rnu \partial +} \left[ \frac{ ( \rnu \partial )^2 }{ \varepsilon^2 } - 1 \right] + 2 \frac{ \ralpha_\partial }{ \rnu^2 \partial_{\! \! \perp}^2 }, ~ \rmu_{\rnu \partial -} = \ralpha_{\rnu \partial -} \left[ \frac{ ( \rnu \partial )^2 }{ \varepsilon^2 } + 1 \right] .
\end{eqnarray}

\noindent The reciprocal matrix is
\begin{eqnarray}                                                           %52
& & ( M^{- 1} )_{\lambda \mu} = \trmu_P P_{\lambda \mu} + \trmu_\rnu \frac{ \rnu_\lambda \rnu_\mu }{ \rnu^2 } + \trmu_\partial \frac{ \partial_{\! \! \perp \lambda} \partial_{\! \! \perp \mu } }{ \partial^2_{\! \! \perp } } + \trmu_{\rnu \partial + } ( \rnu \partial ) ( \rnu_\lambda \partial_{\! \! \perp \mu } + \partial_{\! \! \perp \lambda } \rnu_\mu) \nonumber \\ & & + \trmu_{\rnu \partial - } ( \rnu_\lambda \partial_{\! \! \perp \mu } - \partial_{\! \! \perp \lambda } \rnu_\mu), \nonumber \\ & & \trmu_P = \frac{1}{\rmu_P}, ~ \trmu_\rnu = \frac{\rmu_\partial}{\Delta}, ~ \trmu_\partial = \frac{\rmu_\rnu}{\Delta}, ~ \trmu_{\rnu \partial + } = - \frac{\rmu_{\rnu \partial +}}{\Delta}, ~ \trmu_{\rnu \partial - } = - \frac{\rmu_{\rnu \partial -}}{\Delta}, \nonumber \\ & & \Delta = \rmu_\rnu \rmu_\partial - \rnu^2 \partial_{\! \! \perp}^2 ( \rnu \partial )^2 \rmu^2_{\rnu \partial +} + \rnu^2 \partial_{\! \! \perp}^2 \rmu^2_{\rnu \partial -} = \left( \frac{ \ralpha_\rnu }{ \varepsilon^2 } + \frac{ \ralpha_\partial }{ \rnu^2 \partial_{\! \! \perp}^2 } \right) \left[ - \rnu^4 \frac{ \partial^2 }{ \partial_{\! \! \perp}^2 } \right. \nonumber \\ & & \left. + \ralpha_\partial \frac{ ( \rnu \partial )^4 }{ \varepsilon^2 } + \ralpha_\rnu \rnu^2 \partial_{\! \! \perp}^2 \right] + 4 \frac{ ( \rnu \partial )^2 }{ \varepsilon^2 } \left[ \ralpha_\rnu \ralpha_\partial - \ralpha_\rnu \ralpha_{\rnu \partial +} \rnu^2 \partial_{\! \! \perp}^2 - \ralpha_\partial \ralpha_{\rnu \partial +} ( \rnu \partial )^2 \right] \nonumber \\ & & - \ralpha_{\rnu \partial +}^2 \rnu^2 \partial_{\! \! \perp}^2 ( \rnu \partial )^2 \left[ \frac{ ( \rnu \partial )^2 }{ \varepsilon^2 } - 1 \right]^2 + \ralpha_{\rnu \partial -}^2 \rnu^2 \partial_{\! \! \perp}^2 \left[ \frac{ ( \rnu \partial )^2 }{ \varepsilon^2 } + 1 \right]^2 .
\end{eqnarray}

\noindent By using the freedom to choose $\ralpha_{\lambda \mu}$, we can try to make $( M^{- 1} )_{\lambda \mu}$ possibly well-defined. An example of such a choice looks as follows,
\begin{equation}\label{a=cep2}                                             %53
\ralpha_P = c_0 \varepsilon^2 , \ralpha_\rnu = \ralpha_\partial = c_1 \varepsilon^2 \partial^2 , \ralpha_{\rnu \partial +} = \frac{c_1 + c_2}{ \rnu^2 } \varepsilon^2 , \ralpha_{\rnu \partial -} = c_3 \varepsilon^2 , c_j > 0 , j = 0 - 2 ,
\end{equation}

\noindent $c_j$, $j = 0 - 3$, are constants. Then the denominators $\rmu_P$, $\Delta$ in the expression for $M^{- 1}$ with $\rnu^\lambda = (1, 0, 0, 0)$ in the momentum representation look like
\begin{eqnarray}                                                           %54
& & \Delta = c_1 \left( p^2 \right)^2 \left( 1 + \frac{ \varepsilon^2 }{ \bp^2 } \right) \left( \frac{1}{ \bp^2 } + c_1 p_0^4 + c_1 \varepsilon^2 \bp^2 \right) + 4 c_1 c_2 \varepsilon^2 p_0^2 \left( p^2 \right)^2 \nonumber \\ & & + ( c_1 + c_2 )^2 p_0^2 \bp^2 \left( p_0^2 + \varepsilon^2 \right)^2 + c_3^2 \bp^2 \left( p_0^2 - \varepsilon^2 \right)^2 , ~ \rmu_P = - \frac{1}{ \bp^2 } - c_0 p_0^2 + c_0 \varepsilon^2 .
\end{eqnarray}

\noindent $\Delta$ is positive and can vanish only when $p \to 0$. Further, in the discrete theory, the quantity written as $( \bp^2 )^{- 1}$ in $\rmu_P$ is limited below by a quantity of the order of $b_{\rm s}^2$, where $b_{\rm s}$ is the lattice step in the spatial coordinates. Therefore, for a sufficiently small $\varepsilon$, $\rmu_P$ is strictly negative. Thus, $M^{- 1}$ is practically well-defined. This is important for possible generalizations of the formalism when the metric in the bilinear gauge-violating term is substituted by a nonlinear metric function and vertices containing $M^{- 1}$ arise in perturbation theory. The dependence of $\ralpha_\rnu$, $\ralpha_\partial$ on $\partial$ (\ref{a=cep2}) worsens the UV behaviour of the $\ralpha_{\lambda \mu}$ part of the graviton propagator (\ref{G}). However, again, the discrete underlying formulation limits the short distance behavior; the subsequent passage to the limit $\varepsilon \to 0$, $\ralpha_{\lambda \mu} = O( \varepsilon^2 ) \to 0$ eliminates this part; only the fact of its existence is essential.

\section{Conclusion}

Thus, one can define a prescription for the synchronous graviton propagator similar to the principal value prescription for negative powers of $p_0$ (the "matrix" principal value). It has the following properties.

1) This expression is indeed a propagator for a certain gauge and is obtained by adding to the action a gauge-violating term of general form $\int w_{\lambda \mu} F^{\lambda \mu \sigma \tau} w_{\sigma \tau} \d^4 x $ ($F^{\lambda \mu \sigma \tau}$ is a $10 \times 10$ operator matrix, a function of $\partial$).

2) This term reduces to the more particular form $ \sim \int ( {\rm O }^{\lambda \mu}_\rho w_{\lambda \mu} ) \Lambda^{\rho \kappa} ( {\rm O }^{\sigma \tau}_\kappa w_{\sigma \tau} ) \d^4 x $ with some operators ${\rm O }^{\lambda \mu}_\rho$ and $\Lambda^{\rho \kappa}$. This allows the ghost contribution to be determined by subjecting ${\rm O }^{\lambda \mu}_\rho w_{\lambda \mu}$ to the gauge transformation.

3) The effective ghost action turns out to be of order $O( \varepsilon^2 )$ at $\varepsilon \to 0$ and can be disregarded in the limit $\varepsilon \to 0$ (the subtlety lies in the need for an intermediate regularization by introducing a finite temporal interval $T$ depending on $\varepsilon$, as we consider in \cite{khat1}, so it is essential that the contribution be of order $O( \varepsilon^2 )$ and not just $O( \varepsilon )$).

In general terms, applying this recipe really amounts to taking the principal value of negative powers of $p_0$ (plus corrections that vanish when $\varepsilon \to 0$), which then behave like distributions. For a given diagram we have a finite result of integration over $\d p_0$ at $\varepsilon \to 0$ for almost all other momenta. Good computational properties of such a prescription are due to the fact that the propagator is a sum of (two) terms, each of which has poles only in the upper or only in the lower half-plane of the complex $p_0$, as mentioned in the paragraphs with (\ref{int-(p0+ie)^(-j)...+(+to-)}) and (\ref{vpG=}).

\section*{Acknowledgments}

The present work was supported by the Ministry of Education and Science of the Russian Federation.

\end{document}